\documentclass[conference,a4paper]{IEEEtran}

\usepackage{amssymb, amsmath, amsthm}
\usepackage{graphicx}
\usepackage{cite}

\long\def\symbolfootnote[#1]#2{\begingroup
\def\thefootnote{\fnsymbol{footnote}}\footnote[#1]{#2}\endgroup}

\title{Network Coding Based on Byte-wise Circular Shift and Integer Addition}
\author{\begin{tabular}{cc}
Kenneth W. Shum & Hanxu Hou \\
School of Science and Engineering & School of Electrical Engineering \& Intelligentization \\
The Chinese University of Hong Kong (Shenzhen) & Dongguang University of Technology \\
Shenzhen, China. & Dongguang, China.
 \end{tabular}}

\begin{document}

\maketitle

\begin{abstract}
A novel implementation of a special class of Galois ring, in which the multiplication can be realized by a cyclic convolution, is applied to the construction of network codes. The primitive operations involved are byte-wise shifts and integer additions modulo a power of 2. Both of them can be executed efficiently in microprocessors.  An illustration of how to apply this idea to array code is given at the end of the paper. \symbolfootnote[0]{
This work was partially supported by the
National Natural Science Foundation of China (No. 61701115)}
\end{abstract}

\section{Introduction}

Most of the research works in linear network codes employ finite field as alphabet. It is shown in \cite{LYC03} that the capacity for multicasting a single-source in an acyclic network can be achieved by linear network codes, provided that the field size is sufficiently large. However, large field size incurs large computational complexity. Many practical implementations of network codes employ binary field \cite{XORintheair, LeTehrani}, at the cost of loosing optimality in terms of capacity.

As finite field with odd size is not amenable to computer, finite field with size equal to some power of 2 are preferred in general. The addition of two field elements can be easily done in computer by bit-wise XOR. On the other hand, the computation of field multiplication is more complicated. For field with relatively small size, such as 256,  multiplication can be done by table-lookup method and Zech logarithm (also known as Jacobi logarithm)~\cite[p.79]{Lidl}. However, for very large field size, it would consume too much memory and thus become infeasible. One way to realize finite field arithmetic in a computer is to first fix an irreducible polynomial, and then represent a finite field element by a polynomial. The multiplication is performed by first multiplying two polynomials and then reducing the product by the chosen irreducible polynomial. The multiplication of two field elements can be computed in $O(m^2)$ time, when the field size is $2^m$. When $m$ is large, this introduces a large decoding delay. Furthermore, an intermediate node in a network may be a router with limited processing power and memory. Design of network codes with low computational complexity for real-time applications is a challenging task. 

One approach in lowering the computational complexity of network code implementation is to use bit-wise shift operations,  so that the costly field multiplications can be avoided. The data are first divided into bit strings of fixed length. Then we mix and combine the information stored in the bit strings by cyclic shifts and bit-wise XOR. Because bit-wise cyclic shift and XOR operations are available in most microprocessors, this is an attractive choice for implementing network coding schemes with computational constraint. Indeed, this idea has already been exploited in a series of works  \cite{Xiao08, BASIC, TSLYL, STLYL}. Xiao {\em et al.} first considered cyclic bit-wise shift in the design of network codes for combinatorial networks~\cite{Xiao08}. The BASIC codes in~\cite{BASIC} aim at reducing the computational cost of regenerating codes for distributed storage systems. Linear network coding with cyclic shifts for general single-source acyclic networks is treated in~\cite{TSLYL,STLYL}. We remark that similar methodology is adopted by Blaum and Roth in the context of array codes~\cite{BlaumRoth}.

Besides the XOR operation, another basic arithmetic operation is the addition of unsigned integers. A byte can be regarded as an integer between 0 and 255. Addition is effectively mod-256 addition in the ring of residues $\mathbb{Z}_{256}$. Addition of two integers between 0 and 255 with carry can be done by a single instruction in a microprocessor. It usually takes one clock cycle to execute an addition of integers. The speed is as fast as the speed of performing an XOR operation.  In this paper, we investigate the possibility of byte-wise shifts and integer additions in the design of network codes. The resulting network codes can be easily implemented as in the works mentioned in the last paragraph. 

This paper is organized as follows. We first consider a multi-source network in Section~\ref{sec:example} and demonstrate some potential benefit of employing the ring $\mathbb{Z}_{256}$ instead of a field. An algebraic framework is presented in Section~\ref{sec:algebra}. We give a detailed design example about MDS array code in Section~\ref{sec:arraycode}.

\section{A Motivating Example}
\label{sec:example}

\begin{figure}
\centering
\includegraphics[scale=0.7]{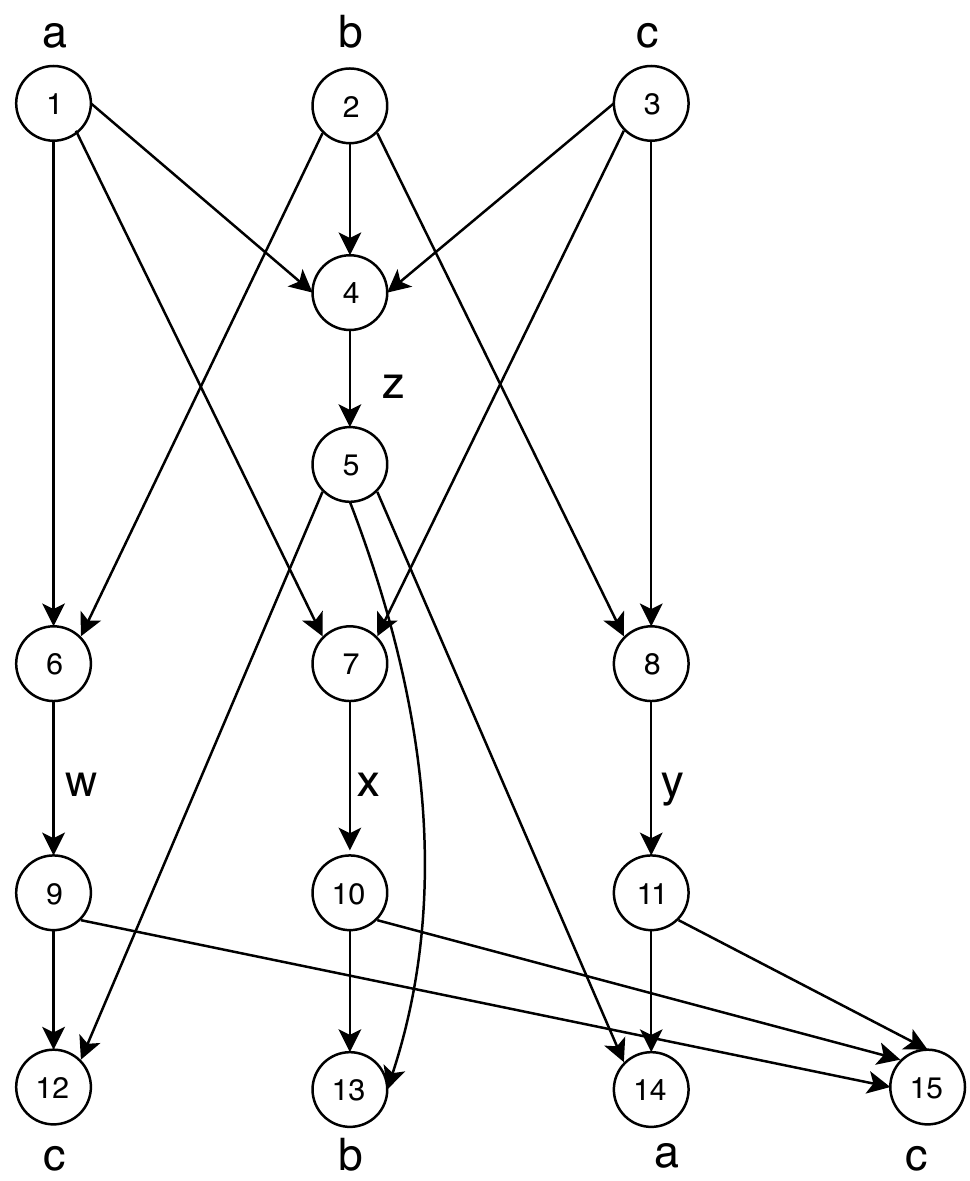}
\caption{A mutli-source network from \cite{DFZ06}.}
\label{figure:DFZ}
\end{figure}

We consider a multi-source network presented in~\cite{DFZ06}. This network has a capacity-achieving solution if the base field has odd characteristic, but any linear network code over a finite field with characteristic 2 will not work. This example is an acyclic network consisting of 15 nodes. The topology of the network is illustrated in Fig.~\ref{figure:DFZ}. Nodes 1, 2 and 3 are the source nodes, and they have symbols $a$, $b$ and $c$ respectively. Nodes 12, 13, 14 and 15 are the sink nodes, and they require symbols $c$, $b$, $a$ and $c$ respectively. It is assumed that we can only send one symbol across each link. In Fig.~\ref{figure:DFZ}, if a node has only one incoming edge, or no incoming edge, then this node can only performing direct forwarding. We do not label the corresponding out-going edges, because the symbol sent from this node is the same as the received symbol.

There are coding opportunity at nodes 4, 6, 7 and 8. If the base field is a finite field with odd characteristic, a solution is to set
\begin{align}
z &= a+b+c \label{coding_example1} \\
w &= a+b \\
x &= a+c \\
y &= b+c, \label{coding_example4}  
\end{align}
where $z$, $w$, $x$ and $y$ are the symbols emitted from nodes 4, 6, 7 and 8, respectively. Upon receiving all the incoming symbols, node 12 can decode $c$ by subtracting $w$ from $z$, node 13 can decode $b$ by subtracting $x$ from $z$, and node 14 can decode $a$ by subtracting $y$ from $z$. Node 15 can solve the following system of linear equations
\begin{equation}
\begin{bmatrix}
1 & 1 & 0 \\
1 & 0 & 1 \\
0 & 1 & 1 \\
\end{bmatrix}
\begin{bmatrix} a \\ b \\ c \end{bmatrix} =
\begin{bmatrix} w \\ x \\ y \end{bmatrix} .
\label{example}
\end{equation}
The determinant of the $3\times 3$ matrix in the previous equation is equal to $-2$. When the alphabet is a finite field with odd characteristic, then node 15 can recover all the source symbols, and in particular, obtain the required symbol $c$. It is proved in~\cite{DFZ06} that this multi-source network is solvable if and only if the alphabet size is odd.

When the base field is a finite field with even characteristic, the $3\times 3$ matrix in the previous paragraph is singular, and we cannot uniquely decode symbol $c$ from symbols $w$, $x$ and $y$. Nevertheless, there is a linear network coding solution over the ring of residues $\mathbb{Z}_{2^m}$, where $m$ is a positive integer, that can get arbitrarily close to the capacity of the network as $m$ tends to infinity. We take $\mathbb{Z}_{2^m}$ as the alphabet, with addition and subtraction performed modulo $2^m$. The source symbol $a$, $b$ and $c$ can assume any value between $0$ and $2^{m-1}-1$. Hence, each source symbol only contains $m-1$ information bits, and there is a 1-bit overhead in each transmission. The symbols $z$, $w$, $x$ and $y$ in the intermediate nodes are computed as in the previous paragraph, with the addition operator replaced by mod-$2^m$ addition, i.e.,
\begin{align*}
z &= a+b+c \bmod 2^m \\
w &= a+b \bmod 2^m\\
x &= a+c \bmod 2^m\\
y &= b+c \bmod 2^m. 
\end{align*}
The decoding in nodes 12, 13 and 14 are the same as described in the previous paragraph. Node 15 computes 
$$x+y-w = 2c \bmod 2^m.$$
Since $c$ is restricted to the range between $0$ and $2^{m-1}-1$, we can solve for the value of $c$ uniquely. We thus see that even though $\mathbb{Z}_{2^m}$ is an alphabet with even cardinality, approximate solution is possible.

Suppose that we use $GF(2^m)$ as the alphabet, instead of $\mathbb{Z}_{2^m}$, and carry out network coding in the intermediate nodes according to Equations \eqref{coding_example1} to \eqref{coding_example4}.
Because the rank of the matrix in Equation~\eqref{example} is equal to 2, there are thus $2^m$ possible solutions in total and they are equally likely. Hence, $c$ can take any value in $GF(2^m)$, and node 15 cannot obtain any information about $c$ in the information-theoretic sense.

This example demonstrates that using the ring $\mathbb{Z}_{2^m}$ can be advantageous in compare to using the finite field $GF(2^m)$, even though their size are identical. When $m=8$, 16, or 32, the addition in $\mathbb{Z}_{2^m}$ is the same as the addition of two unsigned integers with 8, 16, or 32 bits, respectively, and this can be done easily by a single command in a microprocessor. In the next section, we provide a framework on how to use the arithmetic of unsigned integer to construct linear network codes.

\section{Algebraic Formulation}
\label{sec:algebra}

In this paper, a ring $R$ means a commutative ring with identity.   An element $a$ in $R$ is said to be a {\em unit} if there is an element $b \in R$ such that $ab$ is the identity element. A unit is also called an {\em invertible} element. The smallest positive integer $k$ such that $ka$ is equal to the zero element is called the {\em characteristic} of~$R$. Such a positive integer certainly exists when $R$ has finite cardinality. A ring is called a {\em field} if all nonzero elements are invertible. 

We recall some definitions from algebra~\cite[Chapter 3]{Hungerford}. We denote the zero element (resp. the identity element) of $R$ by $0_R$ (resp. $1_R$). An element $a$ in $R$ is called a {\em zero divisor} if we can find a non-zero element $b \in R$ such that $ab=0_R$.  A {\em subring} of $R$ is a subset of $R$ that is also a ring under the addition and multiplication operations of~$R$. We remark that the identity element of the subring $S$ needs not be the same as the identity element of~$R$.   A nonzero element $a \in R$ is {\em irreducible} if (i) $a$ is not a unit, and (ii) $a=bc$ for some $b,c\in R$ implies that $b$ or $c$ is a unit. Two elements $a$ and $b$ in a ring $R$ are said to be {\em coprime} if the ideal generated by $a$ and $b$ is the same as the whole ring~$R$. Equivalently, $a$ and $b$ are coprime if and only if there exist two other elements, say $\lambda$ and $\mu$ in $R$, such that
$$
a \lambda + b \mu = 1_R.
$$

Given a positive integer $n$, the integers $0,1,\ldots, n-1$ form a ring under mod-$n$ addition and multiplication.  In particular, for any positive integer $m$, the ring of residues mod $2^m$, denoted by $\mathbb{Z}_{2^m}$, consists of integers $0,1,2,\ldots, 2^m-1$. All even integers in $\mathbb{Z}_{2^m}$ are zero divisor and they form a maximal ideal. All odd integers in $\mathbb{Z}_{2^m}$ are invertible. 

For any positive integer $p$, the polynomial $x^p-1$ in $\mathbb{Z}_n[x]$ can be factorized as 
\begin{equation}
x^p-1 = (x-1)(x^{p-1} + x^{p-2} + \cdots + x + 1). \label{eq:xp-1}
\end{equation}
In the sequel we will denote the polynomial $x^{p-1} + x^{p-2} + \cdots + x + 1$ by $M_p(x)$. 

 Suppose that $p$ is an odd number. The two polynomials $x-1$ and $M_p(x)=x^{p-1}+x^{p-2}+\cdots + 1$ are coprime, when they are regarded as polynomials in $\mathbb{Z}_{2^m}[x]$. It is because
$$
(x-1)(x^{p-2}+2x^{p-3} + \cdots  (p-2)x
 + p-1) - M_p(x) = -p.
$$
Since $p$ is invertible in $\mathbb{Z}_{2^m}$ when $p$ is odd, we have
$$
(x-1) \alpha(x) + M_p(x) \beta(x) = 1,
$$
where $\alpha(x) = - p^{-1} (x^{p-2} + 2x^{p-3}+\cdots + p-1)$, $\beta(x) = p^{-1}$,
and $p^{-1}$ is the multiplicative inverse of $p$ in $\mathbb{Z}_{2^m}$.

We will fix a prime $p$ such that the multiplicative order of 2 mod $p$ is equal to $p-1$, i.e., $2^i \neq 1 \bmod p$ for all $i\in\{1,2,\ldots, p-2\}$, but $2^{p-1} = 1 \bmod p$. Such a prime number $p$ is said to be a prime with primitive root 2. For a prime $p$ with primitive root 2, the polynomial $M_p(x)$ is irreducible in the polynomial ring $\mathbb{Z}_2[x]$~\cite[Thm 4.1.1]{HuffmanPless}.   For example, $M_p(x)$ is irreducible in $\mathbb{Z}_2[x]$ when $p= 3, 5, 11$ or $13$. The same assumption is also adopted in~\cite{TSLYL, BASIC,Xiao08}. According to the Artin's conjecture~\cite[p.248]{Guy}, there are infinitely many such primes. A list of primes with primitive root 2 can be found at~\cite{OEIS}.

Because the factorization in \eqref{eq:xp-1} is valid over $\mathbb{Z}_{2^m}[x]$ for any positive integer $m$, $M_p(x)$ is a basic irreducible polynomial in $\mathbb{Z}_{2^m}[x]$. The quotient ring $\mathbb{Z}_{2^m}[x] / (M_p(x))$ is isomorphic to the Galois ring with characteristic $2^m$ and size $2^{m(p-1)}$~\cite{ZXWan}. In the following, we use the symbol $GR(2^m, p-1)$ to denote the quotient ring $\mathbb{Z}_{2^m}[x] / (M_p(x))$.

Next, we illustrate an embedding of  $GR(2^m, p-1)$ in $\mathbb{Z}_{2^m}[x]/(x^p-1)$. There are $2^{mp}$ equivalence class in $\mathbb{Z}_{2^m}[x]/(x^p-1)$, and each equivalence class can be represented by a polynomial in the form 
\begin{equation} 
a_{p-1}x^{p-1} +a_{p-2}x^{p-2} + \cdots + a_0, \label{eq:a}
\end{equation}
where $a_i$'s are coefficients chosen from $\mathbb{Z}_{2^m}$. We will identify an equivalence class in $\mathbb{Z}_{2^m}[x]/(x^p-1)$ with a polynomial with degree less than or equal to $p-1$ as in \eqref{eq:a}. With a slight abuse of language, an element in $\mathbb{Z}_{2^m}[x]/(x^p-1)$ will be called a polynomial. We define a mapping  
$$\Phi: \mathbb{Z}_{2^m}[x]/(x^p-1) \rightarrow \mathbb{Z}_{2^m}[x]/(x-1) \oplus \mathbb{Z}_{2^m}/(M_p(x)).$$
Given an equivalence class in $\mathbb{Z}_{2^m}[x]/(x^p-1)$, we choose a representative $a(x)$ in $\mathbb{Z}_{2^m}[x]$ and define $\Phi$
by
$$\Phi(a(x)) := (a(x) \bmod (x-1), a(x) \bmod M_p(x)).
$$
We can verify that $\Phi$ is well-defined and does not depend on the choice of representative. Furthermore, $\Phi$ is an isomorphism by the Chinese remainder theorem~\cite[Thm 2.25]{Hungerford}. The inverse of $\Phi$ is given by
\begin{align*}
\Phi^{-1}(u, v(x)) &= p^{-1} [ u M_p(x)  \\
&\  - v(x) (x-1) ( x^{p-2} + 2x^{p-3} + \cdots + p-1)],
\end{align*}
where $u$ takes value in $\mathbb{Z}_{2^m}$ and $v(x) \in \mathbb{Z}_{2^m}[x]$ is a polynomial with degree less than or equal to $p-2$. Finally, we define a injection from $\mathbb{Z}_{2^m}[x]/(M_p(x))$ to $\mathbb{Z}_{2^m}[x]/(x^p-1)$ by
$$
\Phi^{-1}(0, v(x)).
$$

{\bf Definition.} We denote the image of $\Phi^{-1}(0,\cdot)$, which consists of polynomial $v(x)$ in $\mathbb{Z}_{2^m}[x]$ with degree less than or equal to $p-2$ with $v(1) = 0 \bmod 2^m$, by $R(2^m,p)$. This is a subring of $\mathbb{Z}_{2^m}[x]/(x^p-1)$ that is isomorphic to $GR(2^m,p-1)$. 

\smallskip

{\em Example.} We illustrate the arithmetic of $R(4,3)$, which contains polynomials in the form
$$
a_2 x^2 + a_1 x + a_0.
$$
The coefficients $a_0, a_1, a_2$ are in $\mathbb{Z}_4$ satisfying $a_0+a_1+a_2 = 0 \bmod 4$. Addition of two polynomials in $R(4,3)$ is computed by term-wise mod-4 addition. If $a_2x^2+a_1x+a_0$ and $b_2x^2+b_1x+b_0$ are polynomials in $R(4,3)$, then their sum is equal to
$$
( a_2+b_2) x^2 + (a_1+b_1) x + (a_0+b_0).
$$
The product of $a_2x^2+a_1x+a_0$ and $b_2x^2+b_1x+b_0$ is computed by a cyclic convolution,
\begin{align*}
(a_2x^2+a_1x+a_0)(b_2x^2+b_1x+b_0) \bmod (x^3 - 1) .
\end{align*}
The multiplication of $R(4,3)$ is given in Table~\ref{table:42}.  Each entry in the table consists of three ternary digits $a_2a_1a_0$, and it represents the polynomial $a_2x^2 + a_1x+ a_0$. There are 15 nonzero elements in $R(4,3)$. The identity element is 
$$
\Phi^{-1}(0,1) = 3^{-1}[0\cdot M_3(x)-(x-1)(x+2)] = x^2+x+2,
$$
represented by $112$ in Table~\ref{table:42}. The nonzero elements can be divided into 5 groups. The multiplication table is partitioned into twenty five $3\times 3$ blocks, and each block is a circulant matrix.

\begin{table*}[ht]
\caption{Multiplication table of the nonzero elements in $R(4,3)$.}
\label{table:42}
\begin{center}
\begin{tabular}{|c|ccc|ccc|ccc|ccc|ccc|} \hline
$\cdot$  &112&211&121 &310&031&103 &130&013&301 &332&233&323 &022&202&220\\ \hline \hline 
     112 &112&211&121 &310&031&103 &130&013&301 &332&233&323 &022&202&220\\
     211 &211&121&112 &031&103&310 &013&301&130 &233&323&332 &202&220&022\\
     121 &121&112&211 &103&310&031 &301&130&013 &323&332&233 &220&022&202\\ \hline
		 310 &310&031&103 &112&211&121 &332&233&323 &130&013&301 &022&202&220\\
		 031 &031&103&310 &211&121&112 &233&323&332 &013&301&130 &202&220&022\\ 
		 103 &103&310&031 &121&112&211 &323&332&233 &301&130&013 &220&022&202\\ \hline 
		 130 &130&013&301 &332&233&323 &112&211&121 &310&031&103 &022&202&220 \\
		 013 &013&301&130 &233&323&332 &211&121&112 &031&103&310 &202&220&022\\
		 301 &301&130&013 &323&332&233 &121&112&211 &103&310&031 &220&022&202\\ \hline
		 332 &332&233&323 &130&013&301 &310&031&103 &112&211&121 &022&202&220 \\
		 233 &233&323&332 &013&301&130 &031&103&310 &211&121&112 &202&220&022 \\
		 323 &323&332&233 &301&130&013 &103&310&031 &121&112&211 &220&022&202\\ \hline
		 022 &022&202&220 &022&202&220 &022&202&220 &022&202&220 &000&000&000 \\
		 202 &202&220&022 &202&220&022 &202&220&022 &202&220&022 &000&000&000 \\ 
		 220 &220&022&202 &220&022&202 &220&022&202 &220&022&202 &000&000&000 \\ \hline
\end{tabular}
\end{center}
\end{table*}

\section{Application to the Construction of Array Code}
\label{sec:arraycode}

Despite the abstract algebra involved in the framework provided in the previous section, the implementation can be straightforward. In this section, we illustrate the idea through the design of an array code. Consider a storage system consisting of 6 hard disks. The first four disks store the information bits, while the last two disks store parity-check bits. It is required that the original data can be recovered from any four disks.

We divide the data into chunks of 16 bytes. Within each chunk, we name the 16 bytes by $a_i, b_i, c_i$ and $d_i$ for $i=0,1,2,3$. Disk 1 stores bytes $a_0$ to $a_3$, disk 2 stores bytes $b_0$ to $b_3$, disk 3 stores bytes $c_0$ to $c_3$, and disk 4 stores bytes $d_0$ to $d_3$. Table~\ref{table:arraycode} illustrates the encoding and placement of the information bits. Each entry in Table~\ref{table:arraycode} corresponds to one byte, and take values between 0 and 255. 

The last row in Table~\ref{table:arraycode} is auxiliary.  The values in the last row are chosen such that the sum of the 5 entries in a column is a multiple of 256. The data in the last row is not stored in the storage system. We add the auxiliary row in order to visualize the symmetry of the encoding function.

We take $p$ to be the prime number 5, and let $m=8$. Each column in Table~\ref{table:arraycode} is represented by a polynomial in $R(256,5)$. The first entry in each column is the constant term, while the last entry located in the auxiliary row is the term with degree 4. We let $a(x)$, $b(x)$, $c(x)$, and $d(x)$ be the polynomials associated with disk 1, disk 2, disk 3 and disk 4, respectively. Disk 5 stores the horizontal parity-check bits, represented by the polynomial
$$
e(x) = a(x)+b(x)+c(x)+d(x).
$$
The parity-check bits in disk 6 are computed by adding some cyclically shifted version of the the data in disks 1 to 4. In the ring $R(256,5)$, multiplying a polynomial by
\begin{align*}
s(x) &= \Phi^{-1}(0,x) \mod (x^5-1) \\
&= 51x^4+51x^3+51x^2+52x+51
\end{align*}
is the same as cyclically shifting the coefficients of the polynomial. Thus, the polynomial $s(x)$ defined above acts like a cyclic shift operator. The coded data in disk 6 is computed by
$$
f(x) = a(x) + s(x) b(x) + s(x)^2 c(x) + s(x)^3 d(x).
$$
We note that all the additions are mod-256 additions. The parity-check data are tabulated in the last two columns in Table~\ref{table:arraycode}.

\begin{table*}[ht]
\caption{A $4 \times 6$ Array Code}
\label{table:arraycode}
\begin{center}
\begin{tabular}{|c|c|c|c|c|c|} \hline
Disk 1& Disk 2& Disk 3& Disk 4& Disk 5& Disk 6 \\ \hline \hline
$a_0$ & $b_0$ & $c_0$ & $d_0$ & $a_0+b_0+c_0+d_0$ & $a_0+b_1+c_2+d_3$\\
$a_1$ & $b_1$ & $c_1$ & $d_1$ & $a_1+b_1+c_1+d_1$ & $a_1+b_2+c_3+d_4$\\
$a_2$ & $b_2$ & $c_2$ & $d_2$ & $a_2+b_2+c_2+d_2$ & $a_2+b_3+c_4+d_0$\\
$a_3$ & $b_3$ & $c_3$ & $d_3$ & $a_3+b_3+c_3+d_3$ & $a_3+b_4+c_0+d_1$\\ \hline
$a_4$ & $b_4$ & $c_4$ & $d_4$ & $a_4+b_4+c_4+d_4$ & $a_4+b_0+c_1+d_2$\\ \hline 
\end{tabular}
\end{center}
\end{table*}

In terms of polynomials, the encoding function can be represented by matrix multiplication
$$
\begin{bmatrix} a(x) & b(x) & c(x) & d(x) \end{bmatrix}
\begin{bmatrix}
1 & 0 &0 &0 & 1 & 1 \\
0 & 1 &0 &0 & 1 & s(x) \\
0 & 0 &1 &0 & 1 & s(x)^2 \\
0 & 0 &0 &1 & 1 & s(x)^3
\end{bmatrix}.
$$

In the followings, we show that the original data can be recovered from any four disks. The data in disks 1 to 4 are precisely the information data. It is obvious that we can recover the original data from disks 1 to~4. 

Next, suppose that we have the data from disks 1 to 3 and the data from disk 5. Since $a_i$'s, $b_i$'s and $c_i$'s are known, we can subtract them from the coded data in disk 5. After the subtraction, we get the remaining data symbol $d_i$'s. Using similar procedure, we can download data from any three hard disks among disks 1 to 4, and one of the two parity-check disks, in order to decode the original data.

It remains to show that we can rebuild the original data from any two information disks and the two parity-check disks. To simplify presentation, suppose that we access the data stored in disks 3, 4, 5 and 6. Disks 3 and 4 contain eight information bytes $c_0,\ldots, c_3$ and $d_0 , \ldots, d_3$. We can subtract them from the coded data $e(x)$ and $f(x)$ in disks 5 and disk~6, respectively. Let
\begin{align*}
p_1(x) &:= e(x) - c(x) - d(x) \\
&= a(x) + b(x),  \\
p_2(x) &:= f(x) - s^2(x) c(x) - s(x)^3 d(x) \\
&= a(x) + s(x) b(x).
\end{align*}
The two polynomials $p_1(x)$ and $p_2(x)$ are some combinations of $a(x)$ and $b(x)$,
$$
\begin{bmatrix} p_1(x) & p_2(x) \end{bmatrix} =
\begin{bmatrix} a(x) & b(x) \end{bmatrix}
\begin{bmatrix}
1 & 1 \\ 1 & s(x) 
\end{bmatrix}
$$
We can solve this linear system by post-multiplying both sides of the above equation by a $2\times 2$ matrix $\begin{bmatrix} s(x) & -1 \\ -1 & 1\end{bmatrix}$. This yields
$$
\begin{bmatrix} p_1(x) & p_2(x) \end{bmatrix} 
\begin{bmatrix}
s(x) & -1 \\ -1 & 1 
\end{bmatrix}
= (s(x) - 1)
\begin{bmatrix} a(x) & b(x) \end{bmatrix}.
$$
Now we can solve for $a(x)$ and $b(x)$ from
\begin{align*}
(s(x) - 1) a(x) &= q_1(x) := s(x) p_1(x) - p_2(x), \\
(s(x) - 1) b(x) &= q_2(x) := -p_1(x) + p_2(x).
\end{align*}
All arithmetic operations are performed modulo~256.
Because the above two equations are independent from each other, we only consider the decoding of $a(x)$ below. Suppose $q_1(x) = q_{13}x^3 + q_{12}x^2 + q_{11}x + q_{10}$. We have the following system of linear equations
$$
\begin{bmatrix}
-1 & 1 & 0 & 0 \\
0  & -1& 1 & 0 \\
0  & 0 & -1& 1\\
-1 &-1 &-1 & -2
\end{bmatrix}
\begin{bmatrix} a_0 \\ a_1 \\ a_2 \\ a_3 \end{bmatrix} =
\begin{bmatrix} q_{10} \\ q_{11} \\ q_{12} \\ q_{13} \end{bmatrix}.
$$
The determinant of the $4\times 4$ matrix in the above equation is equal to 5, which is invertible in $\mathbb{Z}_{256}$. We can obtain $a_2$ from
$$
q_{10} + 2 q_{11} - 2 q_{12} - q_{13} = 5 a_2 \bmod 256,
$$
or
$$
a_2 = 5^{-1}(q_{10} + 2 q_{11} - 2 q_{12} - q_{13}) \bmod 256,
$$
where $5^{-1}$ is the multiplicative inverse of $5$ in $\mathbb{Z}_{256}$. Once $a_2$ is known, we can obtain $a_0$, $a_1$ and $a_3$ by solving
\begin{align*}
a_3 &= q_{13} + a_2 \bmod 256, \\
a_1 &= -q_{11} + a_2 \bmod 256, \\
a_0 &= -q_{10} + a_1 \bmod 256.
\end{align*}

This completes the decoding of $a_0$, $a_1$, $a_2$ and $a_3$. The computation of $b_0$, $b_1$, $b_2$ an $b_3$ can be done in a similar fashion.

\section{Conclusion}
In this paper, we describe a simple way to implement a special class of Galois ring. It only involves byte-wise shift and mod-$2^m$ addition. When the exponent $m$ is equal to 8, 16, 32, or 64, the mod-$2^m$ addition of integers can be implemented by a microprocessor instruction in one clock cycle. Byte-wise shift is also easy to implement, because we can use a pointer to access the memory and increment the pointer. We do not need to actually modify and shift the memory content. The new way of implementing the arithmetic of Galois ring may find other applications in codes over ring.




\end{document}